\documentclass{IEEEcsmag}

\usepackage[colorlinks,urlcolor=blue,linkcolor=blue,citecolor=blue]{hyperref}
\usepackage[hyphenbreaks]{breakurl}
\usepackage{upmath,color}
\usepackage{xcolor}
\usepackage{xspace}
\usepackage{subcaption}
\usepackage{pifont}
\usepackage{times}
\usepackage{graphbox}
\usepackage{graphicx}
\usepackage{epstopdf}
\usepackage[htt]{hyphenat}
\usepackage{setspace}
\usepackage{relsize}
\usepackage{textcomp}
\usepackage{circledsteps}
\usepackage{url}
\usepackage[htt]{hyphenat}

\usepackage{balance}

\usepackage{amssymb}
\usepackage[framemethod=tikz]{mdframed}
\usepackage[numbers,sort&compress]{natbib}
\tikzset{
    vertical align/.style={
        baseline=-.5*(height("$+$")-depth("$+$"))
    }
}

\newcommand{\subsect}[1]{\subsection{#1}}

\usepackage{calc}
\usepackage{amsmath,algorithm,algpseudocode}
\usepackage{cases}
\makeatletter
\renewcommand{\Function}[2]{%
  \csname ALG@cmd@\ALG@L @Function\endcsname{#1}{#2}%
  \def\jayden@currentfunction{#1}%
}
\newcommand{\funclabel}[1]{%
  \@bsphack
  \protected@write\@auxout{}{%
    \string\newlabel{#1}{{\jayden@currentfunction}{\thepage}}%
  }%
  \@esphack
}
\makeatother

\usepackage{pythonhighlight}
\definecolor{codegray}{rgb}{0.5,0.5,0.5}
\lstdefinestyle{pythonStyle}{
  basicstyle=\tiny\ttfamily\footnotesize,
  commentstyle=\color{codegray},
  frame=single,
  language=Python,
  stepnumber=1,
  numbers=left,
  numbersep=5pt,
  numberstyle=\tiny\color{codegray},
  tabsize=2,
  showspaces=false,
  showstringspaces=false,
  mathescape,
  moredelim=**[is][\color{red}]{~}{~},
  moredelim=**[is][\color{blue}]{<}{>},
  moredelim=**[is][\color{orange}]{@}{@},
  literate={\\~}{{\textasciitilde}}1
  {\\<}{{\unichar{"003C}}}1
  {\\>}{{\unichar{"003E}}}1
  {\\@}{{\unichar{"0040}}}1
}

\usepackage[utf8]{inputenc}
\usepackage[T1]{fontenc}
\usepackage{soulutf8}
\usepackage{kotex}
\usepackage{lipsum}
\usepackage[normalem]{ulem}
\makeatletter
\def\SOUL@hlpreamble{%
\setul{\dimexpr\dp\strutbox-2pt}{\dimexpr\ht\strutbox+\dp\strutbox-2pt\relax}
\let\SOUL@stcolor\SOUL@hlcolor
\SOUL@stpreamble
}
\makeatother

\newcommand\khlc[1][yellow]{
  \bgroup
  \markoverwith{\textcolor{#1}{\rule[-.5ex]{1pt}{2.5ex}}}
  \ULon
}

\usepackage{tabularx}
\usepackage{array}
\usepackage{multirow}
\usepackage{booktabs}
\usepackage{colortbl}

\makeatletter
\def\hlinewd#1{%
\noalign{\ifnum0=`}\fi\hrule \@height #1 %
\futurelet\reserved@a\@xhline}
\makeatother

\usepackage[framemethod=tikz]{mdframed}
\usepackage{marginnote}
\usepackage{pdfcomment}
\mdfsetup{
  hidealllines=true,
  innerleftmargin=2pt,
  innerrightmargin=2pt,
  innertopmargin=2pt,
  innerbottommargin=2pt,
  leftmargin=-2pt,
  rightmargin=-2pt,
  skipabove=-2pt,
  skipbelow=-2pt,
  backgroundcolor=blue!20}

\newlength{\markerHeight}
\newlength{\markerMargin}
\newlength{\linespace}
\newlength{\linedepth}

\sbox0{yf}
\definecolor{mylime}{RGB}{205, 220, 57}
\definecolor{mygreen}{RGB}{60, 200, 0}
\definecolor{myblue}{RGB}{0, 51, 204}

\definecolor{reviewerA_bg}{RGB}{253, 255, 117}
\definecolor{reviewerB_bg}{RGB}{199, 253, 163}
\definecolor{reviewerC_bg}{RGB}{168, 254, 237}
\definecolor{reviewerD_bg}{RGB}{170, 222, 255}
\definecolor{reviewerE_bg}{RGB}{221, 170, 255}
\definecolor{reviewerF_bg}{RGB}{253, 186, 255}
\definecolor{reviewerG_bg}{RGB}{255, 186, 186}
\definecolor{reviewerH_bg}{RGB}{255, 214, 186}

\jtitle{This manuscript is an extended version of the original paper accepted by IEEE Micro.}
\pubyear{2025}

\begin{document}


\title{CXL Topology-Aware and Expander-Driven Prefetching: Unlocking SSD Performance}

\author{Dongsuk Oh$^{*}$, Miryeong Kwon$^{*}$, Jiseon Kim$^{*}$, Eunjee Na$^{*}$, Junseok Moon$^{*}$, Hyunkyu Choi$^{*}$, Seonghyeon Jang$^{*}$, Hanjin Choi$^{*}$, Hongjoo Jung$^{*}$, Sangwon Lee$^{*}$, Myoungsoo Jung$^{* \dagger \ddagger }$}
\affil
{
\\
\\$^{*}$Next-Generation Silicon and Research Division, \textbf{Panmnesia, Inc.}, Daejeon, South Korea
\\$^{\dagger}$Advanced Product Engineering Division, \textbf{Panmnesia, Inc.}, Seoul, South Korea
\\$^\ddagger$KAIST, Daejeon, South Korea}


\begin{abstract}
\looseness=-1
Integrating compute express link (CXL) with SSDs allows scalable access to large memory but has slower speeds than DRAMs. We present ExPAND, an expander-driven CXL prefetcher that offloads last-level cache (LLC) prefetching from host CPU to CXL-SSDs. ExPAND uses a heterogeneous prediction algorithm for prefetching and ensures data consistency with CXL.mem's back-invalidation. We examine prefetch timeliness for accurate latency estimation. ExPAND, being aware of CXL multi-tiered switching, provides end-to-end latency for each CXL-SSD and precise prefetch timeliness estimations. Our method reduces CXL-SSD reliance and enables direct host cache access for most data. ExPAND enhances graph application performance and SPEC CPU's performance by 9.0$\times$ and 14.7$\times$, respectively, surpassing CXL-SSD pools with diverse prefetching strategies.
\end{abstract}

\maketitle

\label{sec:introduction}
\chapteri{C}ompute Express Link (CXL) is emerging as a key interface for enabling memory disaggregation, where memory resources are decoupled from computing servers to provide scalable access to large-capacity memory \cite{gouk2023memory, ahn2022enabling, sun2023demystifying}. This shift is particularly relevant as storage class memory (SCM) technologies, such as PRAM \cite{oh2006full}, Z-NAND \cite{cheong2018flash}, and XL-Flash \cite{kioxia2022xlflash}, offer significant capacity advantages compared to DRAM. SCM's ability to store large datasets with byte-addressable access makes it an attractive candidate for new memory systems. 
As a result, both industry and academia are exploring the potential of byte-addressable solid-state drives (SSDs) that leverage the CXL protocol to combine SCM's memory semantics with scalable interconnects. 
For example, one approach integrates CXL into Optane SSDs to extend memory hierarchies, while several proof-of-concepts (PoCs) are building CXL-SSDs using advanced flash memory technologies like Z-NAND and XL-Flash \cite{jung2022hello, yang2023overcoming, kwon2023cache}.

Despite the potential of CXL-SSDs for addressing the increasing demand for memory capacity, the underlying SCM technologies remain significantly slower than DRAM. PRAM, for instance, has been shown to exhibit access latencies up to 7$\times$ higher than DRAM \cite{zhang2015study}, while Z-NAND and XL-Flash introduce latencies that are approximately 30$\times$ slower \cite{cheong2018flash}. To mitigate these, many designs incorporate SSD-side DRAM buffers as internal caches, mimicking the architecture of high-performance NVMe storage equipped with substantial internal DRAM \cite{lee2009advances, kim2018autossd, tavakkol2018mqsim}. 
These buffers are effective in reducing write latency but are insufficient to address the long read latencies caused by slow SCM backend media.

Addressing read latency in CXL-SSDs requires a departure from traditional SSD design principles. 
Unlike conventional block devices managed by file systems \cite{rajimwale2009block, bez2023access}, CXL-SSDs directly serve memory requests using load/store operations, bypassing the host-side storage stack. This fundamental shift necessitates understanding the execution patterns of host applications and managing the CPU cache hierarchy to align with the unique performance characteristics of CXL-SSDs. However, current SSD technologies have largely been designed to handle block-level requests, leaving them ill-equipped to address the challenges associated with high-latency memory requests \cite{rajimwale2009block}. As a result, there remains a pressing need for solutions that can bridge this gap, ensuring that CXL-SSDs can fully realize their potential in emerging memory disaggregation architectures.

When CXL-SSDs are integrated into the system memory space as host-managed device memory, existing CPU-side cache prefetching mechanisms can still provide some performance benefits. However, two primary challenges limit the effectiveness of current prefetchers within the cache hierarchy in fully exploiting the advantages of the last-level cache (LLC) for CXL-SSDs: i) \emph{hardware logic size constraints} that hinder the handling of diverse memory access patterns encountered in the CXL memory pooling space, and ii) \emph{latency variations} caused by the differing physical positions of CXL-SSDs within the CXL switch network.

Rule-based cache prefetchers, such as spatial \cite{michaud2016best, bera2019dspatch, bakhshalipour2019bingo} and temporal algorithms \cite{jain2013linearizing, bakhshalipour2018domino, somogyi2009spatio}, often require tens of megabytes of storage—comparable to the size of a typical CPU \emph{last-level cache} (LLC). Due to these high storage requirements, modern CPUs employ simpler prefetching algorithms, such as stream cache prefetchers \cite{intel2011vol3b}. While these are efficient in hardware implementation, they are insufficient to mask the increased latency introduced by CXL-SSDs.

An additional complication arises from the interconnect topology used in CXL-based memory disaggregation. To enable scalable memory expansion, CXL employs a multi-level switch architecture, where each switch level introduces a processing delay. The cumulative latency depends on the position of the target device within the switch network, with deeper levels resulting in higher delays. This latency variation prevents existing prefetchers from retrieving data from CXL-SSDs distributed across the network.
The inability to account for such latency differences further exacerbates the challenge of achieving consistent performance across the disaggregated memory space \cite{sharma2022compute}.

This paper presents an expander-driven CXL prefetcher, \emph{ExPAND}, designed to offload primary LLC prefetching tasks from the host CPU to CXL-SSDs, addressing CPU design area constraints. Implemented within CXL-SSDs, ExPAND employs a heterogeneous machine learning algorithm for address prediction, enabling data prefetching across multiple expander accesses. 
The host-side logic in ExPAND ensures that CXL-SSDs remain aware of the execution semantics of host-side CPUs, while the SSD-side logic maintains data consistency between the LLC and CXL-SSDs using CXL.mem's \textit{back-invalidation} (BI) mechanism. 
This bidirectional collaboration allows user applications to access the majority of data directly at the host, thereby significantly reducing reliance on CXL-SSDs for frequent memory requests.

Accurate estimation of prefetching latency is essential for optimizing the limited capacity of on-chip caches. To address this, we define the concept of \textit{prefetch timeliness}, representing the latency constraints inherent to CXL-based prefetching. ExPAND incorporates a detailed understanding of prefetch timeliness by identifying the CXL network topology and device latencies during PCIe enumeration and device discovery. Using this topology information, ExPAND calculates precise end-to-end latency values for each CXL-SSD within the network and writes these values into the PCIe configuration space of each device. This information enables the offloaded cache prefetching algorithm to determine the optimal timing for transferring data to the host LLC, effectively mitigating the long read latencies caused by the slower backend media of CXL-SSDs.

Our evaluation results show that ExPAND enhances graph application performance and SPEC CPU's performance by 9.0$\times$ and 14.7$\times$, respectively, surpassing CXL-SSD pools with diverse prefetching strategies.


\section{BACKGROUND}
\label{sec:background}


\begin{figure}
  \centering
  \begin{minipage}{.49\linewidth}
      \includegraphics[width=\linewidth]{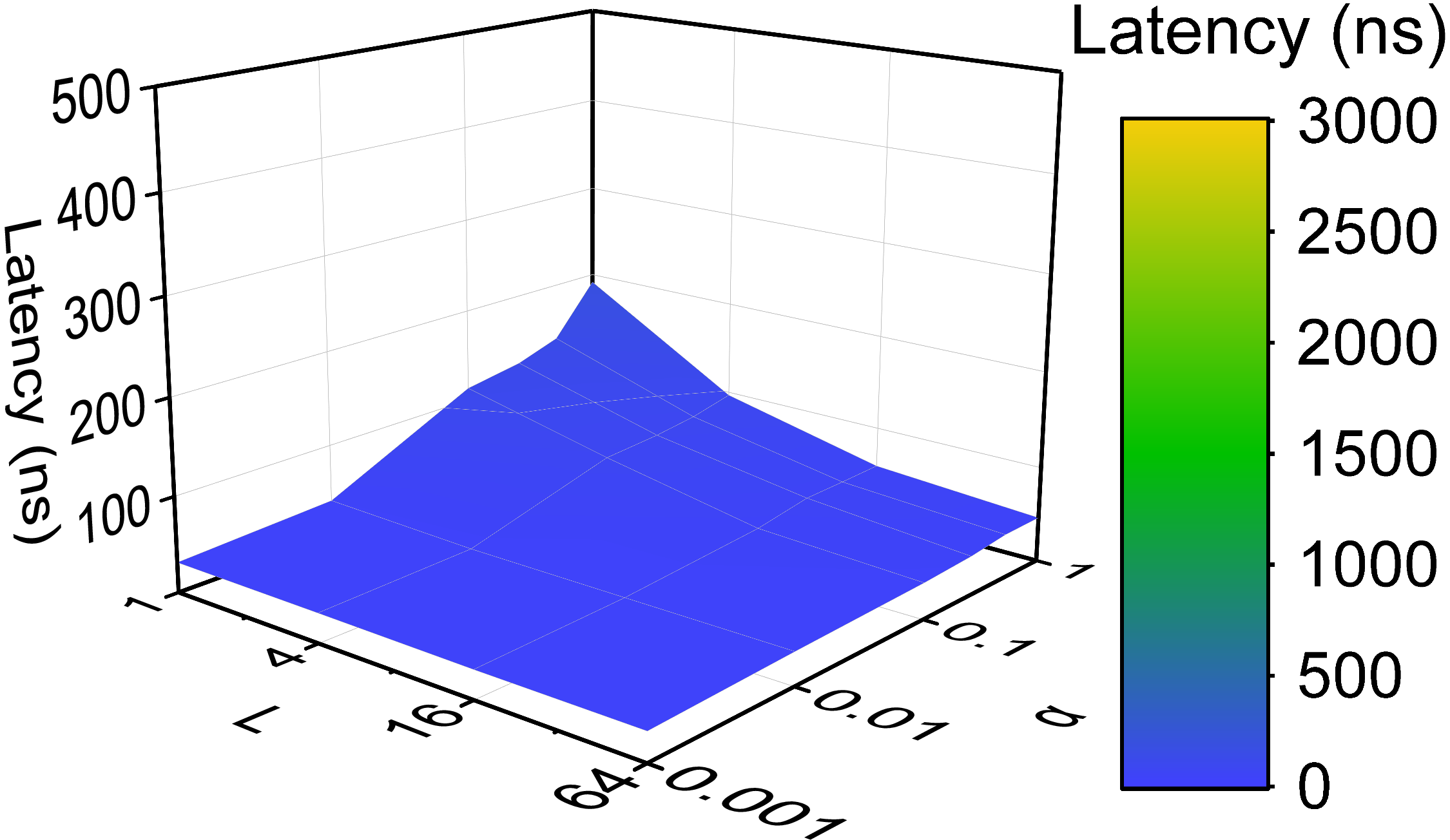}
  \end{minipage}
  \begin{minipage}{.49\linewidth}
      \includegraphics[width=\linewidth]{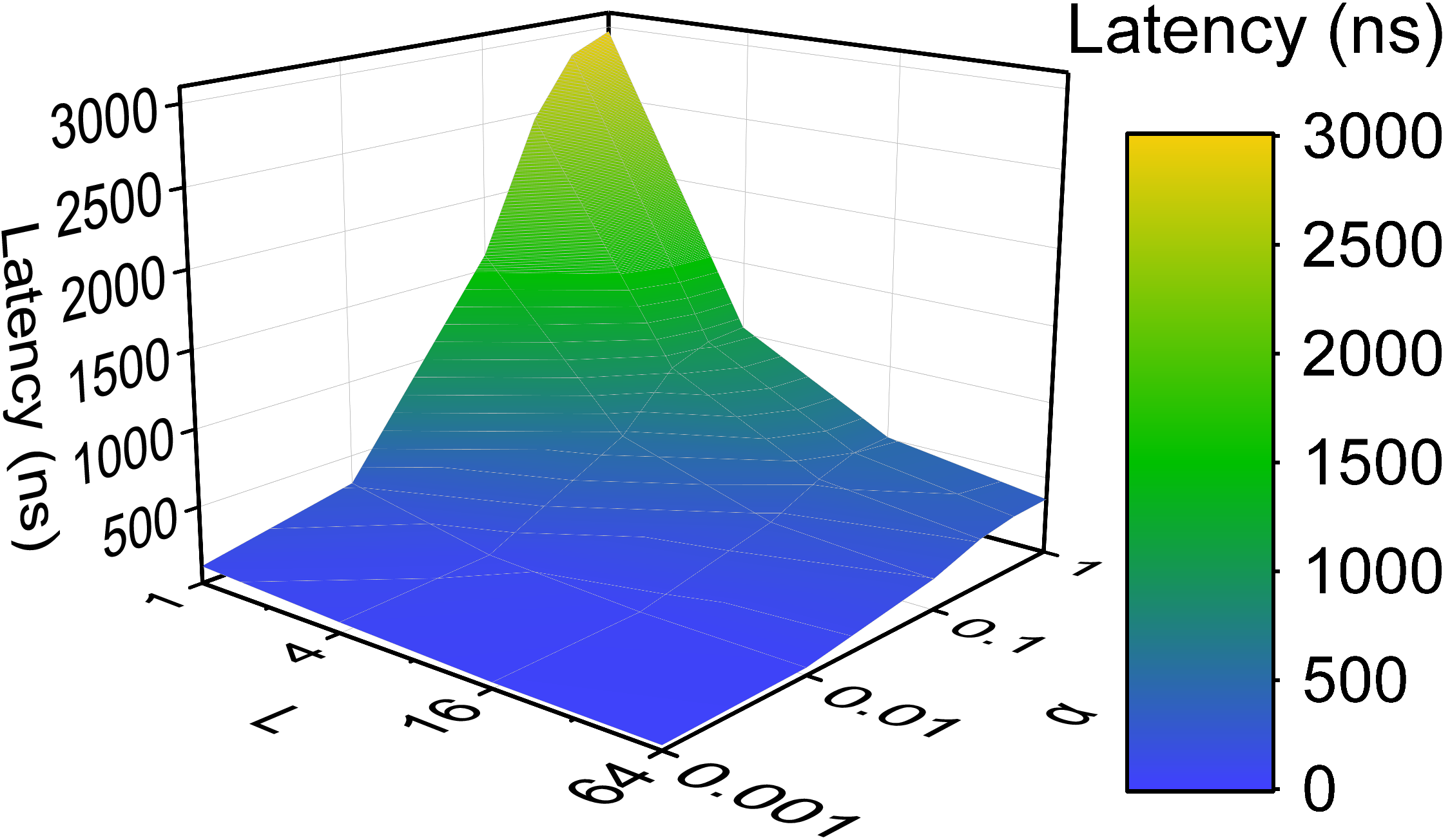}
  \end{minipage}
  \begin{subfigure}{1\linewidth}
      \centering
      \renewcommand*{\arraystretch}{0.3}
      \begin{tabularx}{\textwidth}{
          p{\dimexpr.49\linewidth-2\tabcolsep-1.3333\arrayrulewidth}
          p{\dimexpr.49\linewidth-2\tabcolsep-1.3333\arrayrulewidth}
          }
            \vspace{6pt}\caption{LocalDRAM.} \label{fig:apex1}
          & \vspace{6pt}\caption{CXL-SSD.} \label{fig:apex2}
      \end{tabularx}
  \end{subfigure}

  \caption{Analyzing the impact of locality.} \label{fig:}
\end{figure}

\begin{figure*}
	\centering
	\begin{minipage}{.30\linewidth}
		\includegraphics[width=\linewidth]{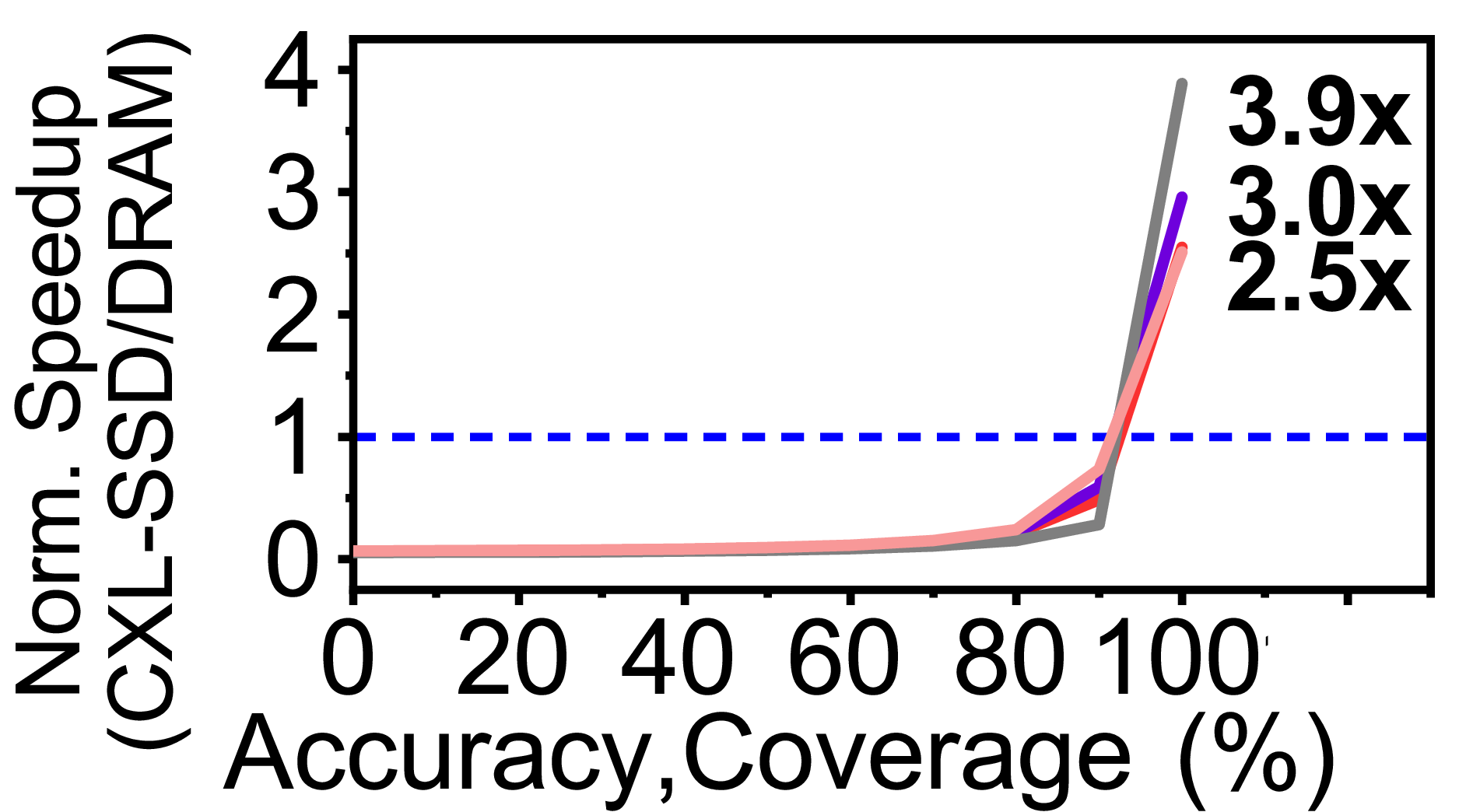}
	\end{minipage}
	\begin{minipage}{.29\linewidth}
		\includegraphics[width=\linewidth]{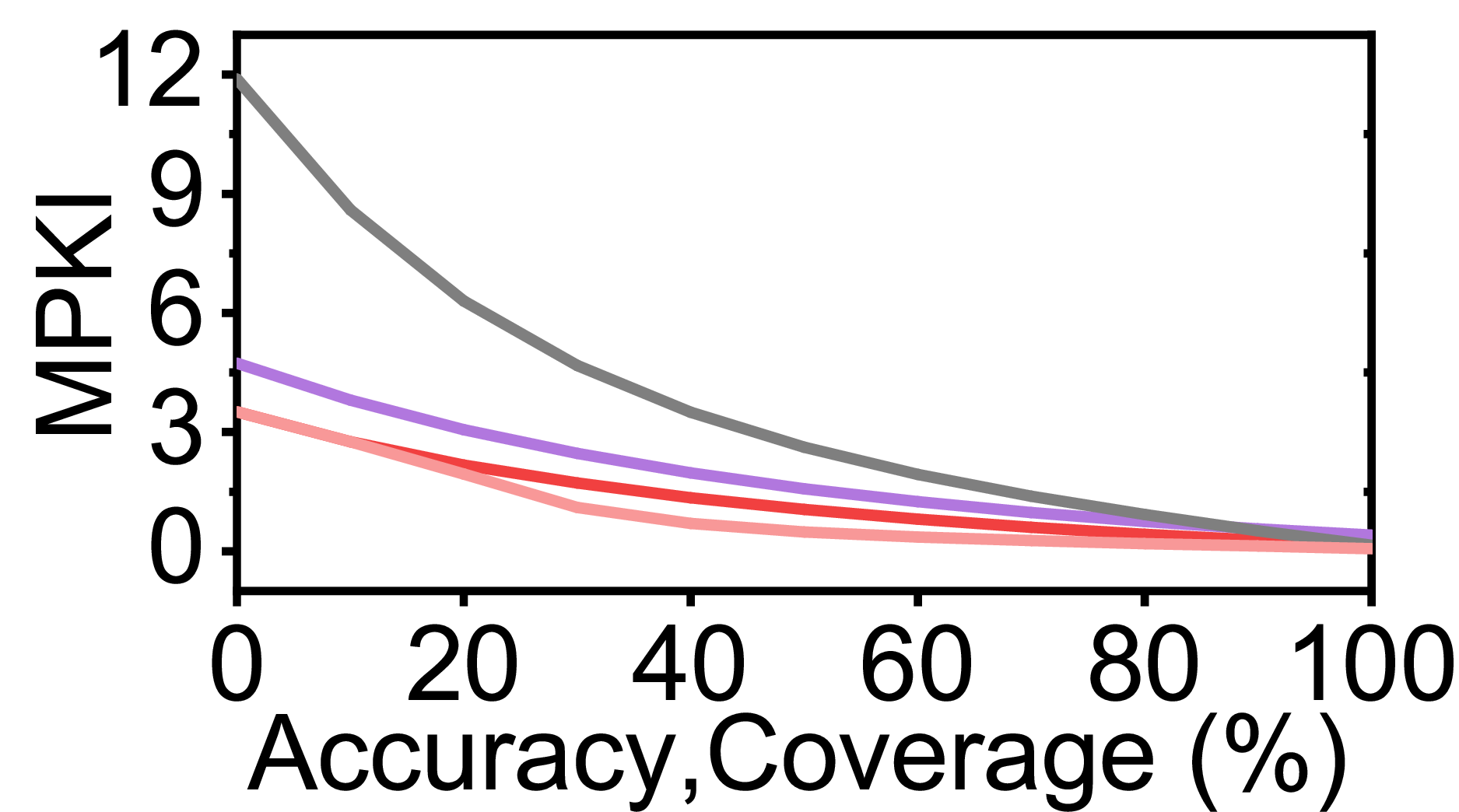}
	\end{minipage}
	\begin{minipage}{.39\linewidth}
		\includegraphics[width=\linewidth]{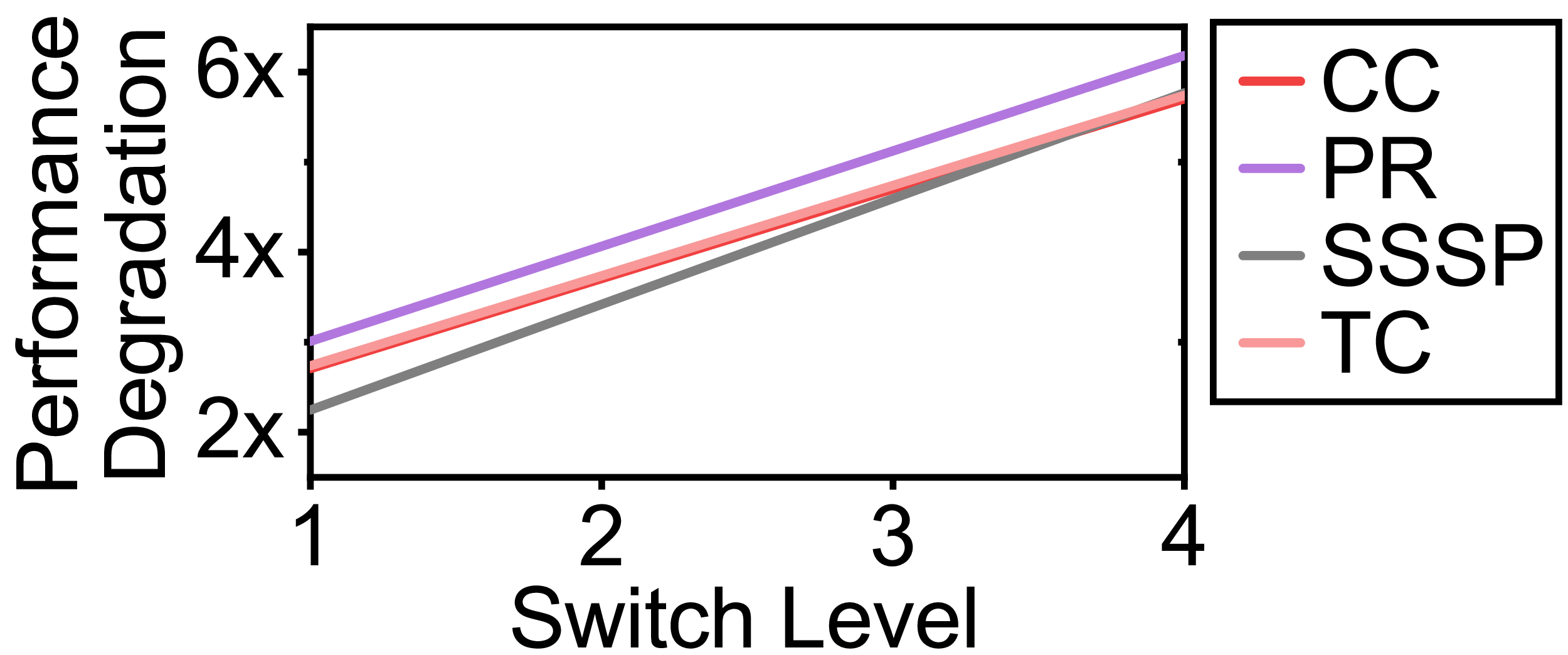}
	\end{minipage}
	\begin{subfigure}{1\linewidth}
		\centering
		\renewcommand*{\arraystretch}{0.3}
		\begin{tabularx}{\textwidth}{
			p{\dimexpr.32\linewidth-2\tabcolsep-1.3333\arrayrulewidth}
			p{\dimexpr.28\linewidth-2\tabcolsep-1.3333\arrayrulewidth}
			p{\dimexpr.40\linewidth-2\tabcolsep-1.3333\arrayrulewidth}
			}
			\vspace{-4pt}\caption{Speedup.} \label{fig:motiv1} &
			\vspace{-4pt}\caption{MPKI.} \label{fig:motiv2} &
			\vspace{-4pt}\caption{Unawareness impact of CXL switches.} \label{fig:motiv3}
		\end{tabularx}
	\end{subfigure}
	
	\caption{CXL-SSD prefetching performance analysis.} \label{fig:motiv}
\end{figure*}

\subsection{Memory Pooling using CXL}
\noindent \textbf{Protocol primary.}
CXL is a cache-coherent interconnect designed for heterogeneous devices, enabling scalable memory expansion. It consists of three sub-protocols: \emph{CXL.io}, \emph{CXL.cache}, and \emph{CXL.mem}. Built on the PCIe physical layer, CXL.io functions as a direct counterpart to the PCIe protocol. CXL.cache facilitates efficient access to host memory for accelerators, while CXL.mem enables hosts to access memory attached to devices across the CXL network. Together, CXL.io and CXL.mem support the connection of multiple memory expanders to create large-scale memory pools. Note that memory expanders can connect to the host system memory without requiring CXL.cache, appearing to CPUs as locally-attached memory. This is possible because CXL allows \emph{endpoint} (EP) devices to be mapped into the cacheable memory space \cite{gouk2023memory, li2023pond, wahlgren2022evaluating, berger2023design}.

\noindent \textbf{Incorporating CXL into storage.}
CXL.mem and CXL.io allow CPU to access memory via load/store instructions, using a CXL message packet called a \emph{flit}. This flit-based communication enables various memory and storage media to be integrated into the CXL pool. SCMs having greater capacity than DRAM is leading to an interest in integrating CXL into block storage, known as \emph{CXL-SSDs}. CXL-SSDs often use large internal DRAM caches to store data ahead of backend SCMs, achieving performance akin to DRAM-based EP expanders. Samsung's PoC employs Z-NAND and a 16GB DRAM cache, claiming an 18$\times$ write latency improvement over NVMe SSDs \cite{samsung2022fms}. Kioxia's PoC uses XL-Flash and a sizable DRAM prefetch buffer, asserting DRAM-like speeds by combining prefetching with hardware compression \cite{kioxia2022xlflash}.


\subsection{Go Beyond Pooling}
\noindent \textbf{Enhanced memory coherence.}
CXL’s flit-based communication decouples memory resources from processor complexes, enabling efficient memory pooling. However, while CXL.cache provides cache coherence, it imposes considerable overhead on EP devices when managing their internal memory. When a CXL-SSD employs CXL.cache to synchronize host-side cache updates, frequent monitoring and approval for memory accesses targeting internal DRAM or backend SCM are required to maintain coherence. To address this, the \emph{back-invalidation} (BI) introduced by CXL 3.0 enables CXL.mem to back-snoop host cache lines \cite{cxl2022spec}. This feature allows EPs, such as CXL-SSDs, to autonomously invalidate host cache lines, reducing dependence on CXL.cache while maintaining coherent memory states.

\noindent \textbf{Multi-tiered switching.}
EP expanders within a pool are interconnected via one or more CXL switches. Each CXL switch consists of \emph{upstream ports} (USPs) and \emph{downstream ports} (DSPs), allowing connections between CPUs and EP expanders. A fabric manager (FM) configures and manages these ports, enabling each host to access its EP expanders through a dedicated data path known as a \emph{virtual hierarchy} (VH). Previously, CXL architectures were restricted to a single switch layer, which limited the capacity of each VH. With the introduction of \emph{multi-tiered switching} in CXL 3.0/3.1, switch ports can now connect to additional switches, significantly increasing the capacity of each VH. This enhancement supports up to 4K devices per VH and accommodates various CXL sub-protocols, greatly improving scalability for resource disaggregation \cite{guo2024cxl}.


\section{MOTIVATION AND CHALLENGES}
\label{sec:motiv}

\subsection{Prefetching Impact Analysis}

\noindent \textbf{Locality impact.}
Figures \ref{fig:apex1} and \ref{fig:apex2} illustrate the latency behavior of locally-attached DRAM (\emph{LocalDRAM}) and CXL-SSDs under varying levels of locality. A global data access benchmark \cite{strohmaier2005apex} is used to synthetically adjust locality levels and evaluate their impact. In this study, $L$ represents the vector length, reflecting spatial locality, while $\alpha$ quantifies temporal locality. An $\alpha$ value of 1 corresponds to minimal locality (purely random access), whereas smaller values indicate greater temporal locality. This approach captures the range of potential performance outcomes and highlights the influence of the cache hierarchy. The test environment
remains consistent with that described in the “EVALUATION”
section, ensuring comparability across all analyses.

In low-locality scenarios ($\alpha$ = 0.1 $\sim$ 1 and $L$ = 4 $\sim$ 16), CXL-SSD performance is, on average, 738\% slower than that of LocalDRAM, a disparity deemed unacceptable in many computing contexts. However, as locality improves, the average latency of CXL-SSDs approaches that of LocalDRAM. At high levels ($\alpha \leq 0.01$ and $L \geq 16$), the performance gap narrows significantly, with CXL-SSD being only 35\% slower than LocalDRAM on average. This improvement is due to data being predominantly retrieved from the LLC rather than backend SCMs. Note that, unfortunately, existing CXL-SSD studies have largely overlooked this behavior, focusing instead on optimizing internal DRAM cache utilization. While effective use of internal DRAM caches is essential for mitigating long-tail latency under low-locality conditions, we argue that improving cache hit rates is equally critical for enhancing user experience with CXL-SSDs.

\noindent \textbf{Prefetching impact.}
The performance impact and effectiveness of prefetching are influenced by two parameters \cite{michaud2016best}: \textit{prefetch accuracy} and \textit{prefetch coverage}. Prefetch accuracy is the proportion of prefetched data actually utilized by the target application, while prefetch coverage measures the fraction of total memory requests served by prefetched data.

Figure \ref{fig:motiv1} shows the latency speedup achieved by applying prefetching techniques in CXL-SSD systems, normalized to the latency of LocalDRAM. The graph demonstrates the relationship between prefetch effectiveness and latency improvements in graph applications. Both parameters were configured with identical values, varying from 0\% to 100\%, to evaluate their combined impact on system behavior. This analysis uses four representative large-scale graph workloads sourced from \cite{leskovec2016snap}, with further details provided in the ``EVALUATION'' section.

The results indicate that CXL-SSD performance is significantly slower than LocalDRAM, with up to 4.5$\times$ slower latency when prefetch effectiveness is below 80\%. However, performance improves substantially as prefetch effectiveness increases. Once prefetch effectiveness exceeds 90\%, the increased cache hit rate reduces the frequency of actual memory accesses, leading to notable latency reductions. While unrealistic, a perfect prefetch allows the CXL-SSD to outperform LocalDRAM, improving latency from 2.5$\times$ to 3.9$\times$.

To better understand the performance improvements by prefetching, we analyze the LLC \textit{misses per kilo instructions} (MPKI) for each workload, as shown in Figure \ref{fig:motiv2}. MPKI increases in the order of CC, TC, PR, and SSSP, and the degree of performance improvement corresponds to this order. 
Specifically, prefetching significantly reduces the memory stall times of SSSP, which is associated with 12 MPKI, resulting in a 3.9$\times$ improvement in application performance.
This improvement occurs because the prefetcher prepares data both accurately and in a timely manner, allowing the CPU to perform tasks without being stalled by memory latency. 
Consequently, improving prefetcher effectiveness not only delivers nonlinear performance gains but also enables speedups that surpass those achievable with conventional DRAM-based systems.
However, realizing such performance improvements requires the design of a prefetcher with high effectiveness, emphasizing both accuracy and coverage.

\subsection{Latency Variation with CXL Switch Topology}
\noindent \textbf{Unawareness impact of CXL switches.}
Even an prefetcher is designed towards having high effectiveness, it cannot fully mitigate performance degradation caused by latency variation in multi-tiered switching environments. 
Specifically, conventional prefetchers fail to account for the additional latency introduced by CXL switches, resulting in data being unavailable when needed by the CPU. 
Consequently, memory requests that should result in cache hits are converted into cache misses.This limitation reduces the effectiveness of the prefetcher in CXL systems, leading to performance degradation.

To evaluate how a multi-tiered CXL switch architecture affects application performance, 
we incrementally increased the number of CXL switch layers from 1 to 4 and measured the resulting performance degradation relative to a baseline system without any switches between the host and CXL-SSD. 
For consistency in analyzing performance trends, we assumed a prefetch effectiveness of 90\%, representing the median value in the rapidly changing effectiveness range in Figure \ref{fig:motiv1}.

Figure \ref{fig:motiv3} shows that the four graph workloads (CC, PR, TC, SSSP) experienced a 2.7$\times$ performance degradation per additional CXL topology switch layer in average. 
The slope of each workload graph reflects the performance degradation, where CC, PR, and TC shows 1.3$\times$ degradation per switch layer, whereas SSSP shows 1.4$\times$ per layer.

The observed performance trends align with the prefetch-induced improvements shown in Figure \ref{fig:motiv1} and
the relationship between performance and MPKI depicted in Figure \ref{fig:motiv2}. 
Workloads that benefit significantly from prefetching typically exhibit reduced execution times, making
them more sensitive to switch-induced latency caused
by cache misses. This suggests that workloads highly reliant
on prefetching are disproportionately affected by the latency
introduced by multi-tiered switches in the CXL topology.

To mitigate the performance losses associated with increasing switch layers, it is essential to design prefetchers that can adapt to latency variations introduced by the CXL
switch hierarchy. Such designs would help preserve the benefits of prefetching even in multi-tiered CXL architectures.

\begin{figure*}
    \centering
    \includegraphics[width=0.49\linewidth]{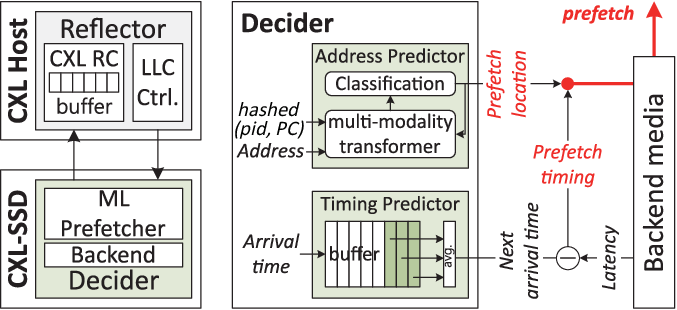}
    \includegraphics[width=0.49\linewidth]{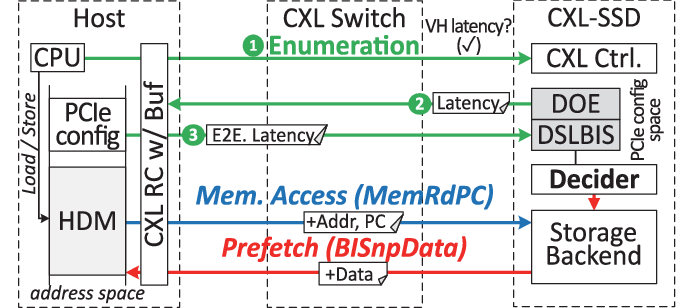}
    \begin{subfigure}{\linewidth}
      \vspace{10pt}
      \centering
      \begin{tabularx}{\textwidth}{
          p{\dimexpr.15\linewidth-1.3\tabcolsep-1.3333\arrayrulewidth}
          p{\dimexpr.345\linewidth-1.3\tabcolsep-1.3333\arrayrulewidth}
          p{\dimexpr.5\linewidth-1.3\tabcolsep-1.3333\arrayrulewidth}
          }
            \vspace{-16pt} \caption{Architecture.} \label{fig:architecture}
          & \vspace{-16pt} \caption{Expander-driven prefetching.} \label{fig:workflow}
          & \vspace{-16pt} \caption{CXL topology-aware prefetch timeliness.} \label{fig:design}
      \end{tabularx}
    \end{subfigure}
    \vspace{-12pt}
    \caption{Overview of ExPAND.} \label{fig:overview}
  \end{figure*}

\section{Expander-Driven Prefetching}
\vspace{3pt}

CPU-side rule-based spatial \cite{michaud2016best, bera2019dspatch, bakhshalipour2019bingo} and temporal \cite{jain2013linearizing, bakhshalipour2018domino, somogyi2009spatio} prefetchers have been adopted in various industrial processors. However, their accuracy is limited (9\% to 76\%) for workloads with large-scale, irregular, or random memory access patterns, insufficient to accelerate CXL-SSD performance to match Local DRAM levels (e.g., 90\% accuracy).
To address these limitations, more advanced prefetching techniques incorporating machine learning (ML) approaches have been proposed \cite{zhang2022fine, zhang2023phases, hashemi2018learning, braun2019understanding, srivastava2020memmap, narayanan2018deepcache, zhang2022transformap, shi2021hierarchical,duong2024tlite,zhang2024dart}. Prefetching involves prediction, making ML-based approaches promising for higher accuracy. While these techniques could achieve the required accuracy threshold, they remain impractical for on-chip CPU implementation due to substantial storage requirements for model computation and metadata management overhead \cite{shi2021hierarchical,duong2024tlite,zhang2024dart}.

On the other hand, techniques to minimize the memory overhead of ML-based prefetching algorithms have also been proposed \mbox{\cite{duong2024tlite,zhang2024dart}}. Unfortunately, these methods either require profiling-based offline training using workload memory traces, thus limiting their effectiveness for accelerating unseen workloads \cite{duong2024tlite}, or utilize knowledge distillation and product quantization techniques to reduce memory requirements, resulting in low accuracy and high training complexity \cite{zhang2024dart}. In addition, existing prefetching algorithms are unaware of multi-level switch architectures, thus unable to account for latency variations inherent in CXL topology.


\subsection{Prefetching Delegation and Collaboration}
This paper introduces an \emph{expander-driven prefetcher} (ExPAND) that delegates cache prefetching decisions to CXL-SSDs, enabling autonomous CPU cache line updates. By shifting decision-making to the EP side, ExPAND leverages the larger form factor and computational capabilities of SSD EPs versus on-chip CPUs. Figure \ref{fig:architecture} presents ExPAND's architecture, which comprises two key components: the \emph{reflector} and the \emph{decider}.

The reflector is implemented on the host-side CXL \emph{root complex} (RC) and LLC controller. Its main role is to provide the decider with essential decision-making inputs like program counter (PC) and switch depth of connected CXL-SSD.
It also communicates the cache prefetching results determined by the decider.
To support this, the reflector uses a small buffer (16 KB) to log cache line updates prefetched by the decider. 
The reflector ensures efficient data delivery by enabling each host's LLC controller in the CXL network to first check the buffer. 
If the required data is present in the CXL RC, the LLC controller directly serves the data from the buffer, avoiding unnecessary traversal through the CXL-SSD pool.

On the other hand, the decider resides in the EP-side CXL-SSD controller, implements a heterogeneous ML prefetcher optimized for irregular memory access patterns \cite{zhang2023phases}. 
Using the provided inputs (PC and memory address), the decider identifies and transfers data to the reflector buffer. 
In addition, the decider records the input data for online refinement of prefetching patterns. 
Detailed explanations of the prefetcher's operation and the interaction mechanisms between the reflector and decider are provided shortly.

\begin{table*}[]
  \begin{minipage}[t]{.31\linewidth}
    \centering
    \resizebox{1\linewidth}{!}{%
      \setlength{\tabcolsep}{0.5pt}
      \setlength\extrarowheight{0.5pt}
      \begin{tabular}{cl}
        \hline
        \textbf{Param.}                & \multicolumn{1}{c}{\textbf{Value}} \\
        \hline
        \textbf{CPU}                   & O3 12cores@3.6GHz, 512-entry ROB  \\
        \textbf{L1 I\$}                & 32KB 2-way, 5-cycle latency        \\
        \textbf{L1 D\$}                & 48KB 2-way, 5-cycle latency        \\
        \textbf{L2\$}                  & 1.25MB 16-way 20-cycle latency     \\
        \hline
        \multirow{2}{*}{\textbf{DRAM}} & tRP = tRCD = tCAS = 22ns,          \\
                                       & 8 Rank, 16 Bank, 2 Channel         \\
        \textbf{PMEM}                  & Intel P5800X                       \\
        \hline
        \textbf{PCIe/CXL}              &  64.0 GT/s (PCIe 6.0), CXL 3.0     \\
        \hline
      \end{tabular}}
    \subcaption{Host configurations.} \label{tab:setup1}
  \end{minipage}%
  \begin{minipage}[t]{.25\linewidth}
    \centering
    \resizebox{1\linewidth}{!}{%
      \setlength{\tabcolsep}{0.5pt}
      \setlength\extrarowheight{0.6pt}
      \begin{tabular}{cl}
        \hline
        \textbf{Param.}                                                       & \multicolumn{1}{c}{\textbf{Value}}     \\
        \hline
        \multirow{2}{*}{\textbf{NAND Flash}}                                  & Samsung 983 ZET, 2TB                    \\
                                                                              & tRd: 3$\mu$s, tWr: 100$\mu$s \\
        \hline
        \multirow{2}{*}{\shortstack{\textbf{Internal} \\ \textbf{DRAM}}}      & tRP = tRCD = 9.1ns,                    \\
                                                                              & tRAS=19ns, size = 1.5GB                \\
        \hline
        \multirow{3}{*}{\shortstack{\textbf{Location} \\ \textbf{Predictor}}} & Attention dim: 64                \\
                                                                              & Modality fusion dim: 128         \\
                                                                              & Transformer dim: 128             \\
        \hline
        \textbf{Timing}                                                       & Buffer entries: 10                     \\
        \hline
      \end{tabular}}
      \subcaption{CXL-SSD configurations.} \label{tab:setup2}
  \end{minipage}%
  \begin{minipage}[t]{0.191\linewidth}
    \resizebox{\textwidth}{!}{%
      \setlength{\tabcolsep}{0.5pt}
      \setlength\extrarowheight{0.6pt}
      \begin{tabular}{lrrr}
        \toprule
        \multirow{2}{*}{\textbf{Workload}}  & \multirow{2}{*}{\shortstack{\textbf{Working} \\ \textbf{set (GB)}}} & \multirow{2}{*}{\,\,\textbf{MPKI}\,\,} & \multirow{2}{*}{\,\shortstack{\textbf{Read} \\ \textbf{ratio}}\,} \\
                        & & & \\ 
        \midrule
        PR              & 82           & 4.13        & 0.01  \\
        SSSP            & 428          & 11.03       & 0.01  \\
        TC              & 31           & 3.13        & 0.01  \\
        bwaves          & 22           & 0.27        & 0.84  \\
        leslie3d        & 41           & 0.45        & 0.52  \\
        lbm             & 22           & 0.28        & 0.03  \\
        libquantum      & 141          & 1.48        & 0.63  \\
        mcf             & 215          & 12.17       & 0.87  \\
        \bottomrule
      \end{tabular}}
    \subcaption{Workloads.} \label{tab:workload}
  \end{minipage}%
  \begin{minipage}[t]{0.255\linewidth}
    \resizebox{\textwidth}{!}{%
    \setlength{\tabcolsep}{0.5pt}
      \begin{tabular}{cccc}
        \toprule
        \multirow{2}{*}{\textbf{Algorithm}} & \multirow{2}{*}{\shortstack{\textbf{Memory} \\ \textbf{overhead}}} & \multirow{2}{*}{\textbf{IOPs}} & \multirow{2}{*}{\textbf{Accuracy}} \\
                       & & & \\ 
        \midrule
        Prior work 1\cite{duong2024tlite} & 64KB        & 56.6K         &  86\% \\
        Prior work 2\cite{zhang2024dart}  & 548.8KB     & 4.9M          &  81\% \\
        Rule 1                            & 4KB         & 768           &  82\% \\
        Rule 2                            & 8KB         & 2304          &  53\% \\
        ML 1                              & 936.8KB     & 11.3M         &  88\% \\
        ML 2                              & 865KB       & 26M           &  89\% \\
        ExPAND                            & 839.2KB     & 10.3M         &  92\% \\
        \bottomrule
      \end{tabular}}
    \subcaption{Prefetch algorithms.} \label{tab:prefetch_algo}
  \end{minipage}
  \vspace{2pt}
  \caption{Evaluation setup.} \label{tab:setup}

\end{table*}

\subsection{Prefetch Address and Timing Speculation}
\label{sec:prefetcherdetials}
As shown in Figure \ref{fig:workflow}, ExPAND's decider incorporates two specialized predictors: an address predictor and a timing predictor, responsible for determining the location and timing of cache prefetch operations, respectively. The address predictor is inspired by a multi-modality transformer model \cite{zhang2023phases}, while the timing predictor employs a simpler rule-based approach.
The address predictor generates a sequence of memory addresses for prefetching by utilizing a transformer as its sequence model. It integrates a multi-modality attention network \cite{jewitt2016introducing} to enhance the analysis of relationships between memory access patterns and PC.
In addition, it monitors changes in application execution behavior using a decision tree classifier \cite{myles2004introduction} to dynamically refine prefetching accuracy.


ExPAND's decision tree classifier is pretrained to categorize memory traces of various applications into 64 categories. For online inference, ExPAND maintains a sliding window containing recent memory addresses and their corresponding PCs, feeding this information to the classifier model. The classifier infers the current window's requests into one of the pretrained 64 categories. If the classifier's inference changes from the previously inferred category, ExPAND records this as a behavior-change event. Such events are then provided as hints, along with memory addresses, to the multi-modality transformer model. By recognizing these behavior-change events, the transformer model achieves more accurate predictions of subsequent addresses. This detection-based feedback allows the transformer model to respond promptly to changes in memory access patterns.

The timing predictor, in contrast, maintains request arrival time information in a small-sized buffer (80B) and estimates future memory request times by averaging historical arrival times within its history window. Accurate prediction requires retaining all past arrival times in the history window. However, when memory requests are served directly by the LLC, relevant information may not reach the timing predictor. To address this, the reflector notifies the decider of cache hit events via CXL.io, enabling the timing predictor to account for future request times even when no direct requests are observed.

The actual prefetch timeliness is determined by combining the timing predictor's results with each device's latency variation. 
A detailed discussion of the prefetch timeliness estimation process is provided in the “CXL Topology-Aware Prefetch Timeliness” section.

\section{CXL CROSS-LAYER INTERSECTION}
\label{sec:topology}

\subsection{CXL Topology-Aware Prefetch Timeliness}
\label{subsec:timeliness}
\noindent \textbf{CXL switch hierarchy discovery.}
The reflector effectively identifies the switch level of CXL-SSDs during PCIe enumeration. 
The host accesses the configuration space of connected devices (using CXL.io) and organizes system buses within their CXL network. During enumeration, buses are segregated upon identifying new devices, with unique number is assigned to each bus. As CXL switch operates as PCIe bridge device with distinct bus number, it allows determining the number of switches between a host CPU and target CXL-SSD. The reflector stores this information on its RC side, which aids in estimating accurate prefetch timeliness, detailed below.

\noindent \textbf{Timeliness speculation.}
Prefetching data too early could contaminate the LLC, reducing its hit ratio, while prefetching too late may delay execution. 
Therefore, pinpointing exact prefetch timeliness is essential. Figure \ref{fig:design} shows this process. Since CXL EPs must manage \emph{data object exchange} (DOE) capability in PCIe configuration space, each CXL-SSD can determine its device latency through DOE. However, this latency cannot directly estimate prefetch timeliness due to variations from multi-tiered switching. During enumeration, the reflector retrieves each CXL-SSD's latency by extracting \emph{device scoped latency and bandwidth information structure} (DSLBIS) from DOE. It then calculates the latency overhead incurred in VH between RC and target CXL-SSD.  The reflector combines this VH latency with the DSLBIS latency and stores the end-to-end latency in the corresponding device's configuration space. Consequently, the decider estimates prefetch timeliness by subtracting the end-to-end latency from the time predicted by its timing predictor.


\begin{figure*}
  \centering
  \begin{minipage}{.49\linewidth}
		\includegraphics[width=\linewidth]{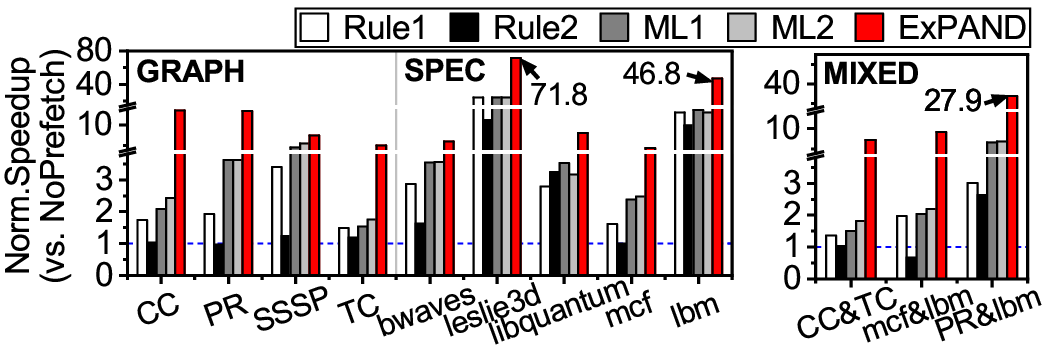}
	\end{minipage}
  \begin{minipage}{.32\linewidth}
		\includegraphics[width=\linewidth]{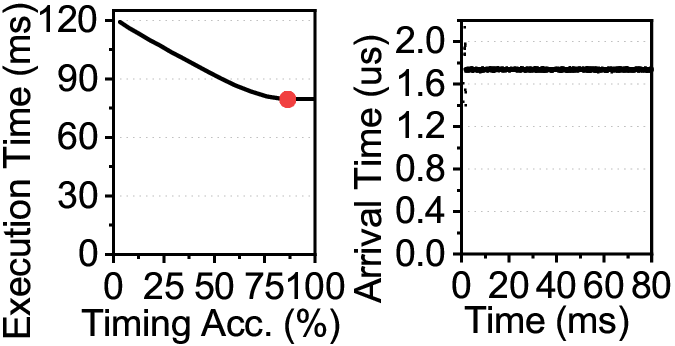}
	\end{minipage}
  \begin{minipage}{.175\linewidth}
		\includegraphics[width=\linewidth]{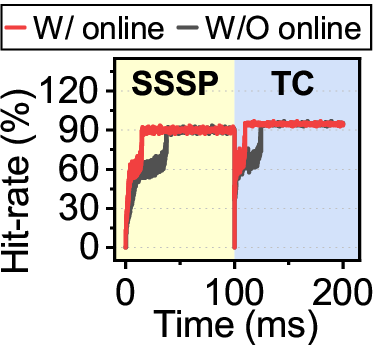}
	\end{minipage}

  \begin{subfigure}{1\linewidth}
      \centering
      \renewcommand*{\arraystretch}{0.3}
      \begin{tabularx}{\textwidth}{
          p{\dimexpr.31\linewidth-2\tabcolsep-1.3333\arrayrulewidth}
          p{\dimexpr.2\linewidth-2\tabcolsep-1.3333\arrayrulewidth}
          p{\dimexpr.16\linewidth-2\tabcolsep-1.3333\arrayrulewidth}
          p{\dimexpr.16\linewidth-2\tabcolsep-1.3333\arrayrulewidth}
          p{\dimexpr.175\linewidth-2\tabcolsep-1.3333\arrayrulewidth}
          }
            \vspace{-4pt}\caption{Speedup compared to NoPrefetch.} \label{fig:eval_over}
          & \vspace{-4pt}\caption{Mixed workload.} \label{fig:eval_mixed}
          & \vspace{-4pt}\caption{Sensitivity.} \label{fig:eval_sensitivity}
          & \vspace{-4pt}\caption{Arrival time.} \label{fig:eval_arrival}
          & \vspace{-4pt}\caption{Online tuning.} \label{fig:eval_timeline}
      \end{tabularx}
      \vspace{1pt}
  \end{subfigure}
  
  \vspace{-8pt}
	\caption{Overall performance.} \label{fig:eval_overall}
\end{figure*}

\subsection{Bidirectional Communication on CXL}
\noindent \textbf{Downward: piggybacking on CXL.mem.}
To accurately predict addresses, timely transmission of PCs and corresponding memory requests is vital. CXL.mem's master-to-subordinate (\emph{M2S}) transactions include request without data (\emph{Req}), request with data (\emph{RwD}), and back-invalidation response (\emph{BIRsp}). Req is primarily for memory read opcode (\emph{MemRd}) without payload, while RwD carries payload for memory write opcode (\emph{MemWr}). RwD allows 13 custom opcodes, enabling an opcode for memory reads with PCs (\emph{MemRdPC}). When a read misses the LLC, the reflector sends an M2S transaction using MemRdPC, including the current PC. Consequently, the target decider can access the memory address and PC in the host's execution environment. Note that BIRsp responds to the CXL-SSD's BI snoop command, discussed shortly.

\noindent \textbf{Upward: leveraging BI.}
When it reaches the time to prefetch (estimated in the ``CXL Topology-Aware Prefetch Timeliness'' section), the decider must update the reflector buffer with data obtained from its address predictor results. However, the existing CXL.mem lacks the capability to update host-side on-chip storage. To address this limitation, we use the CXL.mem subordinate-to-master (S2M) transaction's BISnp. BISnp, similar to CXL.mem's Req, is a non-payload message. We introduce a new BI opcode, termed \emph{BISnpData}, to the S2M transaction message, allowing up to 10 custom opcodes. Using BISnpData, the decider generates a payload accompanying its message, containing data for updating the host. When the reflector detects BISnpData, it awaits the corresponding payload and inserts the prefetched data into its buffer, enabling the LLC controller to fetch it for execution.

\section{EVALUATION}
\label{sec:evaluation}
\noindent \textbf{Methodologies.}
Since no CXL-SSDs are currently available, we use CXL hardware RTL modules in a full system simulation model. We conduct this simulation using gem5 \cite{binkert2011gem5} and SimpleSSD \cite{gouk2018amber}. 
These CXL RTL modules have been validated using a real CXL end-to-end system at the cycle-level \cite{gouk2022direct}. 
Table \ref{tab:setup} provides the main parameters of our simulation. We compare ExPAND, our proposed expander-driven prefetcher, with several modern rule-based and ML-based prefetchers. 
These include a spatial prefetcher \cite{michaud2016best} (\texttt{Rule1}), a temporal prefetcher \cite{jain2013linearizing} (\texttt{Rule2}), an LSTM-based prefetcher \cite{shi2021hierarchical} (\texttt{ML1}), and a transformer-based prefetcher \cite{zhang2022fine} (\texttt{ML2}). We summerize the important characteristics of each prefetching algorithm in Table \ref{tab:prefetch_algo}.


\noindent \textbf{Workload and benchmarks}
To evaluate the effectiveness of ExPAND in diverse graph application scenarios, we employ four widely used algorithms across five datasets. The algorithms, sourced from established graph processing frameworks, include Connected Components (\textit{CC}), PageRank (\textit{PR}), Single Source Shortest Path (\textit{SSSP}), and Triangle Counting (\textit{TC}). These are applied to five datasets: Amazon's product co-purchasing network, Google's web graph, the California road network, the Wikipedia talk network, and the YouTube online social network. We also summarize key characteristics of each workload in Table \ref{tab:workload}.

\subsect{Overall Analysis of Prefetching} We evaluate all five prefetching techniques (\texttt{Rule1}, \texttt{Rule2}, \texttt{ML1}, \texttt{ML2}, and \texttt{ExPAND}) across the graph workloads and SPEC CPU benchmarks. For clarity, we normalize their performance relative to \texttt{NoPrefetch}, representing a CXL-SSD without prefetching.
Figure \ref{fig:eval_over} analyzes the speedup of the five prefetching techniques normalized to \texttt{NoPrefetch}.

\noindent \textbf{Rule-based prefetcher.}
In graph applications, \texttt{Rule1} achieves a performance improvement over \texttt{NoPrefetch} and \texttt{Rule2} by 2$\times$ and 1.8$\times$, respectively. This enhancement occurs because graph applications typically exhibit significant spatial locality, making data accesses easier for prefetchers to predict compared to temporal patterns.
In contrast, performance varies across SPEC CPU benchmarks. Workloads with high MPKI (e.g., mcf) still perform comparably to \texttt{NoPrefetch}. However, workloads exhibiting structured, repetitive memory access patterns achieve an average speedup of 7$\times$.

\noindent \textbf{ML-based prefetcher.}
ML-based prefetchers achieve performance improvements of 1.6$\times$ over rule-based prefetchers and 4.4$\times$ over \texttt{NoPrefetch}. This gain is attributed to their ability to effectively learn complex spatial and temporal locality patterns.
\texttt{ExPAND}, our proposed multi-modality transformer-based prefetcher, further outperforms existing ML-based prefetchers by 2.4$\times$, achieving speedups ranging from 4.3$\times$ up to 71.8$\times$ over \texttt{NoPrefetch}. \texttt{ExPAND} particularly excels in workloads dominated by stencil computations, such as \texttt{bwaves}, \texttt{leslie3d}, and \texttt{lbm}. Stencil computations typically involve referencing neighboring data points across multiple dimensions.


 \begin{figure*}
\begin{minipage}{0.49\linewidth}
	\centering
  \includegraphics[width=1\linewidth]{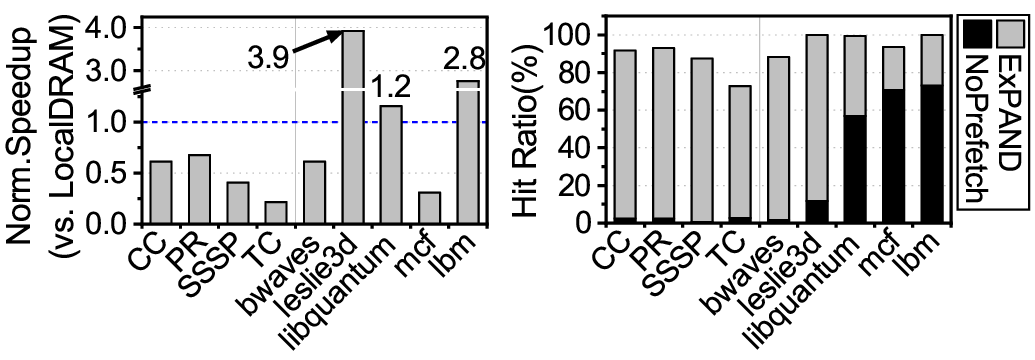}

  \vspace{-14pt}
  \begin{subfigure}[t]{.49\linewidth}
    \caption{Speedup.}
    \label{fig:perf-speedup}
	\end{subfigure}
  \begin{subfigure}[t]{.49\linewidth}
    \caption{LLC hit ratio.}
    \label{fig:llc-hit-ratio}
	\end{subfigure}

  \caption{Performance comparison with local DRAM.} \label{fig:eval_media_switch}
\end{minipage}
\begin{minipage}{0.49\linewidth}
	\centering
  \includegraphics[width=1\linewidth]{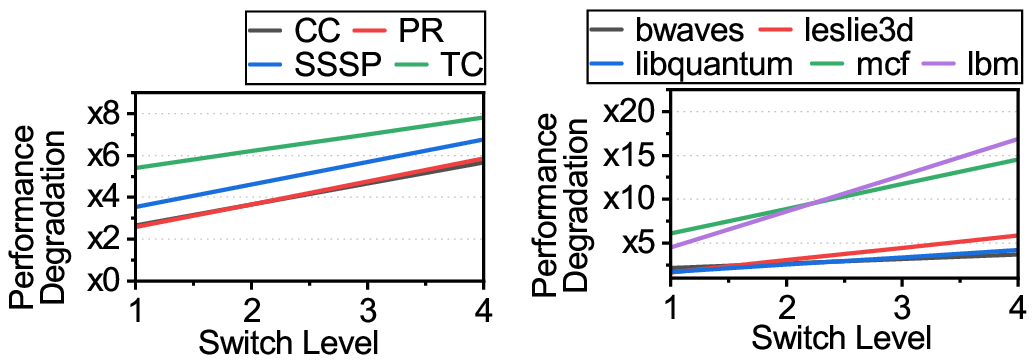}

  \vspace{-14pt}
  \begin{subfigure}[t]{.49\linewidth}
    \caption{Graph workload.}
    \label{fig:timely-graph}
	\end{subfigure}
  \begin{subfigure}[t]{.49\linewidth}
    \caption{SPEC CPU benchmark.}
    \label{fig:timely-spec}
	\end{subfigure}

  \caption{Impacts of timeliness.} \label{fig:eval_media_switch}
\end{minipage}
\end{figure*}



\noindent \textbf{Performance with mixed workloads.}
Figure \ref{fig:eval_mixed} illustrates the execution time results under mixed workload scenarios, where each core simultaneously runs distinct workloads. The performance of existing prefetching algorithms (\texttt{Rule1}, \texttt{ML1}, and \texttt{ML2}) significantly degrades under mixed workloads due to reduced accuracy in predicting subsequent memory addresses resulting from intertwined memory access patterns. An exception is \texttt{Rule2}, which preprocesses memory accesses by grouping addresses with similar values. Consequently, \texttt{Rule2} maintains relatively high performance when workloads with strong spatial locality (e.g., CC \& TC) are combined.

In contrast, \texttt{ExPAND} employs a multi-modality transformer model considering both PC and memory addresses, enabling it to effectively distinguish between memory access patterns even under mixed workloads. Therefore, under mixed workloads, \texttt{ExPAND} outperforms \texttt{Rule1}, \texttt{Rule2}, \texttt{ML1}, and \texttt{ML2} by averages of 7.0$\times$, 10.2$\times$, 3.7$\times$, and 3.5$\times$, respectively.

\noindent \textbf{Model optimizations.}
We further evaluate ExPAND's prefetching optimizations, particularly its timeliness model and online tuning mechanisms. ExPAND's timeliness model considers backend media latency of CXL expansion devices and multi-layer switch latency, enabling efficient utilization of limited LLC resources.

Figure \ref{fig:eval_sensitivity} shows workload performance relative to the accuracy of the timeliness model, evaluated using the TC workload. As shown, improved timeliness accuracy directly correlates with reduced execution time. Low accuracy leads either to early prefetching -- causing prefetched data to be evicted before usage -- or delayed prefetching, resulting in data being unavailable when required. Performance gains begin saturating at around 68\% timeliness accuracy, with marginal improvements beyond 84\%. This saturation occurs because LLC associativity typically mitigates eviction issues arising from minor timing inaccuracies.

ExPAND's timeliness model achieves 90\% accuracy, significantly improving workload performance. Despite being heuristic-based, this high accuracy is achievable since LLC access frequencies remain relatively constant during workload execution. Figure \ref{fig:eval_arrival} illustrates the intervals between LLC accesses during execution of the TC workload. As seen, LLC access frequency remains stable at runtime, influenced primarily by the workload's inherent memory access frequency and randomness, both of which typically remain constant throughout execution.


\begin{figure*}
	\centering
  \begin{minipage}{.49\linewidth}
    \includegraphics[width=1\linewidth]{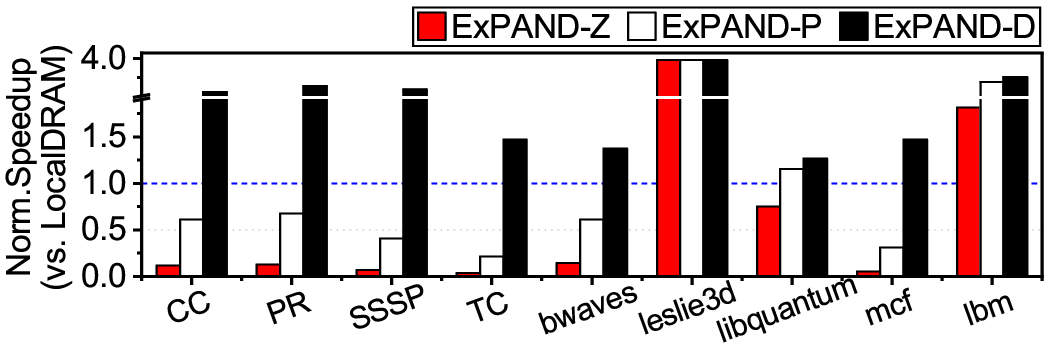}
	\end{minipage}
  \begin{minipage}{.49\linewidth}
    \includegraphics[width=\linewidth]{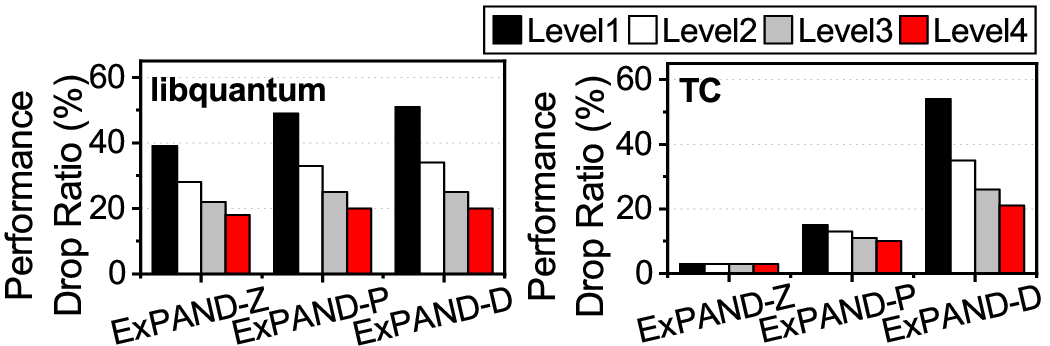}
	\end{minipage}

  \vspace{-5pt}
  \begin{subfigure}{1\linewidth}
    \renewcommand*{\arraystretch}{0.3}
    \begin{tabularx}{\textwidth}{
      p{\dimexpr.50\linewidth-2\tabcolsep-1.3333\arrayrulewidth}
      p{\dimexpr.50\linewidth-2\tabcolsep-1.3333\arrayrulewidth}
      }
      \vspace{-1pt}\caption{Overall performance.} \label{fig:eval_media} &
      \vspace{1pt}\caption{Multi-switch analysis.} \label{fig:eval_media_switch1}
    \end{tabularx}

	\end{subfigure}

  \caption{Impacts of backend media.} \label{fig:eval_media_switch}

  \vspace{-10pt}
\end{figure*}

\noindent \textbf{Performance gain.}
Figure \ref{fig:eval_timeline} analyzes the performance gains from ExPAND's online tuning. The evaluation measures LLC hit rates in scenarios with and without online tuning, focusing on dynamic workload transitions. Two workloads, SSSP and TC, were chosen due to their contrasting memory access patterns -- SSSP has sequential accesses, whereas TC involves large-stride patterns.

The results show that online tuning enables significantly faster recovery of LLC hit rates following behavioral changes in workloads. Without tuning, the transformer model struggles to quickly adapt due to reliance on patterns from a large historical window, making it difficult to detect sudden changes. In contrast, ExPAND's decision tree classifier rapidly identifies changes in workload behavior, immediately notifying the transformer model. Consequently, the transformer adjusts swiftly by prioritizing recent memory accesses.

\subsect{Performance Comparison with Local DRAM.}

In the previous subsection, we analyze the performance improvements from expander-driven prefetching on CXL-SSD. Here, we evaluate its practical effectiveness during application execution. For this analysis, we compare \texttt{ExPAND}, the most effective prefetching technique, against a baseline using only local DRAM (\texttt{LocalDRAM}).


\noindent \textbf{Impacts of prefetching.}
Figure \ref{fig:perf-speedup} shows the normalized performance of \texttt{ExPAND} against \texttt{LocalDRAM} in terms of application execution time. To better understand the performance gap, we also measure the LLC hit ratio, illustrated in Figure \ref{fig:llc-hit-ratio}. 
The \texttt{NoPrefetch} percentage reflects the baseline LLC hit ratio without prefetching, while \texttt{ExPAND} highlights the additional improvements achieved through prefetching. \texttt{ExPAND} delivers a 9.0$\times$ performance improvement over \texttt{NoPrefetch} across four graph workloads. However, compared to the \texttt{LocalDRAM} baseline, it shows a 48\% performance degradation due to the 14\% cache miss rate, which required accessing CXL-SSD. Consequently, application execution is delayed despite \texttt{ExPAND} achieving an 86\% LLC hit rate.
In contrast, SPEC benchmarks such as leslie3d, libquantum, and lbm demonstrate 3.9$\times$, 1.2$\times$, and 2.8$\times$ performance improvements over \texttt{LocalDRAM}, respectively. This improvement is driven by \texttt{ExPAND}'s ability to increase the LLC hit ratio by 46\% on average, achieving hit rates as high as 96\% for theses workloads.

\noindent \textbf{Impacts of timeliness.}
We conduct a sensitivity evaluation by incrementally increasing the VH level of the CXL switch from 1 to 4 to understand the impact of latency variation with topology.
Figure \ref{fig:timely-graph} and Figure \ref{fig:timely-spec} show the impact of increasing switch levels on performance for each workloads. 
The performance trends align with the LLC hit ratio in \ref{fig:llc-hit-ratio}. 
Applications with high \texttt{ExPAND}-driven cache hit rates experience significant performance degradation at switch level 1 due to prefetchers that are unaware of the CXL topology, leading to reduced prefetch effectiveness. 
Conversely, workloads with higher \texttt{NoPrefetch} LLC hit ratios experience steeper degradation as levels increase, as added switch latency affects applications that had benefit from high LLC hit rates.
Graph workloads (Figure \ref{fig:timely-graph}) show consistent degradation, averaging a 1.2$\times$ slowdown. 
SPEC CPU benchmarks (Figure \ref{fig:timely-spec}) exhibit varied slowdowns reflecting different sensitivities to switch latency.

\subsection{Diversity of Backend Media}
To evaluate the impact of backend media on expander-driven prefetching and application execution time, we conducted experiments with Z-NAND, PMEM, and DRAM as backend media options. These configurations are referred to as \texttt{ExPAND-Z}, \texttt{ExPAND-P}, and \texttt{ExPAND-D}, respectively.
\texttt{ExPAND-Z} and \texttt{ExPAND-P} were tested to examine the feasibility of using Z-NAND and PMEM as main memory, while \texttt{ExPAND-D} was analyzed to explore the maximum potential benefits of expander-driven prefeching.


\noindent \textbf{Impacts of prefetching.}
Figure \ref{fig:eval_media} presents the execution times of graph workloads and SPEC benchmarks using memory expanders with different backend media.
\texttt{ExPAND-Z}, which utilizes Z-NAND as its backend (6$\times$ slower than PMEM), exhibits an average of 3$\times$ higher performance degradation compared to \texttt{ExPAND-P}. However, for workloads such as leslie3d and lbm, \texttt{ExPAND-Z} achieves 3.9$\times$ and 1.8$\times$ better performance, respectively, than \texttt{LocalDRAM}. This indicates that, for specific workloads, a PMEM-based memory expansion system can outperform \texttt{LocalDRAM} configurations.

\texttt{ExPAND-D}, leveraging DRAM as its backend, outperforms \texttt{LocalDRAM} across all graph workloads and SPEC benchmarks, with performance improvements ranging from 1.3$\times$ to 3.9$\times$ and an average gain is 1.9$\times$. This shows that as the performance of the
backend media in memory expander improves, the benefits of using memory expander become much greater compared to building a computing system with only local DRAM.


\noindent \textbf{Impacts of timeliness.}
To evaluate the impact of backend media on prefetching timeliness, we perform a switch-level sensitivity analysis using libquantum (highest LLC hit ratio) and TC (lowest LLC hit ratio) as representative workloads (cf. Figure \ref{fig:llc-hit-ratio}). 
For libquantum, the high LLC hit ratio minimizes the impact of backend media latency, making switch latency the dominant factor as shown in Figure \ref{fig:eval_media_switch1}. 
In contrast, TC's low LLC hit ratio amplifies backend media latency, reducing the impact of switch latency. \texttt{ExPAND-Z} and \texttt{ExPAND-P} show 15\% and 3\% degradation at switch level 1, with average reductions of 12\% and 3\% for further levels. \texttt{ExPAND-D}, with its low backend latency, experiences a 54\% drop at switch level 1 and an average of 32\% for additional levels, highlighting its sensitivity to switch latency.
These results emphasize the need for prefetchers to achieve high LLC hit rates and account for CXL topology to mitigate switch latency effects, especially with high-performance backend media.


\section{CONCLUSION}
\label{sec:conclusion}
We propose an expander-driven CXL prefetcher that offloads LLC prefetching to CXL-SSDs, employing a heterogeneous prediction algorithm. ExPAND ensures data consistency and provides precise prefetch timeliness estimates, reducing CXL-SSD reliance and enhancing graph application performance and SPEC CPU's performance by 9.0$\times$ and 14.7$\times$, respectively,
compared to other prefetching strategies.
This work is protected by one or more patents.

\def\refname{REFERENCES}
\vspace*{-20pt}
\renewcommand\refname{\section{REFERENCES}}
\bibliographystyle{unsrtnat}
\balance
\bibliography{reference}

\end{document}